\documentstyle[11pt]{article}
\input amssymb.sty

\title{Epistemological and Ontological \\Paraconsistency in Quantum Mechanics:\\
For and Against Bohrian Philosophy.}

\author{{\sc C. de Ronde}}
\date{}

\begin{document}

\bibliographystyle{plain}
\maketitle

\begin{center}
\begin{small}
Philosophy Institute ``Dr. A. Korn" \\ 
Buenos Aires University, CONICET - Argentina \\
Center Leo Apostel and Foundations of  the Exact Sciences\\
Brussels Free University - Belgium \\
\end{small}
\end{center}

\begin{abstract}
\noindent We interpret the philosophy of Niels Bohr as related to the so called ``linguistic turn'' and consider paraconsistency in the light of the Bohrian notion of complementarity. Following \cite{daCostaKrause06}, Jean-Yves B\'eziau has discussed the seemingly contradictory perspectives found in the quantum mechanical double slit experiment in terms of paraconsistent viewpoints \cite{Beziau12, Beziau14}. This interpretation goes in line with the well known Bohrian Neo-Kantian epistemological account of quantum mechanics. In the present paper, we put forward the idea that one can also consider, within quantum mechanics and departing from the philosophy of the danish physicist, a more radical paraconsistency found within one of the main formal elements of the theory, namely, quantum superpositions. We will argue that, rather than epistemological, the contradictions found within quantum superpositions could be interpreted as ontological contradictions. 
\end{abstract}
\begin{small}

{\em Keywords: quantum superposition, paraconsistent approach, measurement problem.}

\end{small}

\bibliography{pom}

\begin{thebibliography}{10}

\bibitem{ArenhartKrause14a} Arenhart, J. R. and Krause, D., 2014, ``Contradiction, Quantum Mechanics, and the Square of Opposition", {\it Philosophy of Science Archive}, http://philsci-archive.pitt.edu/10582/

\bibitem{ArenhartKrause14b} Arenhart, J. R. and Krause, D., 2014, ``Oppositions in Quantum Mechanics",  in {\it New dimensions of the square of opposition}, Jean-Yves B\'eziau and Katarzyna Gan-Krzywoszynska (Eds.), 337-356, Philosophia Verlag, Munich. {\it Philosophy of Science Archive}, http://philsci-archive.pitt.edu/10583/

\bibitem{ArenhartKrause14c} Arenhart, J. R. and Krause, D., 2014, ``Potentiality and Contradiction in Quantum Mechanics.", in {\it Festschrift honoring J.-Y. Beziau's 50th Birthday}, in press. {\it Philosophy of Science Archive}, http://philsci-archive.pitt.edu/10726/

\bibitem{Bacciagaluppi96} Bacciagaluppi, G., 1996, {\it Topics
in the Modal Interpretation of Quantum Mechanics}, Doctoral
dissertation, University of Cambridge, Cambridge.

\bibitem{Nature13} Bernien, H., Hensen, B., Pfaff, W., Koolstra, G., Blok, M. S., Robledo, L., Taminiau, T. H., Markham, M., Twitchen, D. J., Childress, L. and Hanson, R., 2013, ``Heralded entanglement between solid-state qubits separated by three metres'',  {\it Nature}, {\bf 497}, 86-90.

\bibitem{Bitbol10} Bitbol, M., 2010, ``Reflective Metaphysics:
Understanding Quantum Mechancis from a Kantian Standpoint", {\it
Philosophica}, {\bf 83}, 53-83.

\bibitem{Beziau12}  B\'eziau, J.-Y., 2012, ``The Power of the Hexagon'', {\it Logica Universalis}, {\bf 6}, 1-43.

\bibitem{Beziau14}  B\'eziau, J.-Y., 2014, ``Paraconsistent logic and contradictory viewpoint'', to appear in {\it Revista Brasileira de Filosofia}, 241.

\bibitem{Bohr60} Bohr, N., 1960, {\it The Unity of Human Knowledge},
In {\it Philosophical writings of Neils Bohr}, vol. 3., Ox Bow
Press, Woodbridge.

\bibitem{BokulichCP} Bokulich, A., 2014, ``Bohr's Correspondence Principle'', {\it The Stanford Encyclopedia of Philosophy (Spring 2014 Edition)}, Edward N. Zalta (Ed.), URL: http://plato.stanford.edu/archives/spr2014/entries/bohr-correspondence/.

\bibitem{BokulichBokulich} Bokulich, P., and Bokulich, A., 2005, ``Niels Bohr's Generalization of Classical Mechanics'', {\it Foundations of Physics}, {\bf 35}, 347-371.

\bibitem{Bub97} Bub, J., 1997, {\it Interpreting the
Quantum World}, Cambridge University Press, Cambridge.

\bibitem{PS} Curd, M. and Cover, J. A., 1998, {\it Philosophy of Science. The central
issues}, Norton and Company (Eds.), Cambridge University Press,
Cambridge.

\bibitem{daCostadeRonde13} da Costa, N. and de Ronde, C., 2013, ``The Paraconsistent Logic of Quantum Superpositions", {\it Foundations of Physics}, {\bf 43}, 845-858.

\bibitem{daCostadeRonde14} da Costa, N. and de Ronde, C., 2014, ``The Paraconsistent Approach to Quantum Superpositinos Reloaded", preprint.

\bibitem{daCostaKrause06} da Costa, N. C. A. and Krause, D., 2006, ``The Logic of Complementarity'', In {\it The Age of Alternative Logics: Assessing Philosophy of Logic and Mathematics Today}, J. van bent hem, G. Heinzmann, M. Rebuschi and H. Vesser (Eds.), 103-120, Springer.

\bibitem{daCostaKrauseBueno07} da Costa, N. C. A., Krause, D., and Bueno, O.,
``Paraconsistent logics and paraconsistency'', in {\it Handbook of
the Philosophy of Science (Philosophy of Logic)}, D. Jacquette
editor, Elsevier, 2007, pp. 791-911.

\bibitem{deRonde05} de Ronde, C., 2005, ``Complementary Descriptions (PART 1)", {\it Los Alamos Archive}, arXiv: quant-ph/0507105v1. 

\bibitem{deRonde10} de Ronde, C., 2010, ``For and Against Metaphysics in the Modal Interpretation of Quantum Mechanics", {\it Philosophica}, {\bf 83}, 85-117.

\bibitem{deRondePHD} de Ronde, C., 2011, {\it The Contextual and Modal Character of Quantum Mechanics: A Formal and Philosophical Analysis in the Foundations of
Physics}, PhD dissertation, Utrecht University.

\bibitem{deRonde12} de Ronde, C., 2012, ``La noci\'on de potencialidad en la interpretaci\'on modal de la mec\'anica cu\'antica'', {\it Scientiae Studia}, 2012, {\bf 1}, 137-164.

\bibitem{deRonde13a} de Ronde, C., 2013, ``Quantum Superpositions and Causality: On the Multiple Paths to the Measurement Result", {\it Philosophy of Science Archive}, http://philsci-archive.pitt.edu/10049/

\bibitem{deRonde13b} de Ronde, C., 2013, ``Representing Quantum Superpositions: Powers, Potentia and Potential Effectuations", {\it Philosophy of Science Archive}, http://philsci-archive.pitt.edu/10155/

\bibitem{deRonde14YQM} de Ronde, C., 2014, ``The Problem of Representation and Experience in Quantum Mechanics'', in {\it Probing the Meaning of Quantum Mechanics: Physical, Philosophical and Logical Perspectives},  91-111, D. Aerts, S. Aerts and C. de Ronde (Eds.), World Scientific, Singapore.

\bibitem{deRonde14a} de Ronde, C., 2014, ``A Defense of the Paraconsistent Approach to Quantum Superpositions (Answer to Arenhart and Krause)", {\it Philosophy of Science Archive}, http://philsci-archive.pitt.edu/10613/

\bibitem{deRonde14b} de Ronde, C., 2014, ``Modality, Potentiality and Contradiction\\in Quantum Mechanics", WCP5, Springer, sent. 

\bibitem{RFD14a} de Ronde, C., Freytes, H. and Domenech, G., 2014, ``Interpreting the Modal Kochen-Specker Theorem: Possibility and Many Worlds in Quantum Mechanics'', {\it Studies in History and Philosophy of Modern Physics}, {\bf 45}, pp. 11-18.

\bibitem{RFD14b}  de Ronde, C., Freytes, H. and Domenech, G., 2014, ``Quantum Mechanics and the Interpretation of the Orthomodular Square of Opposition'', in {\it New dimensions of the square of opposition}, Jean-Yves B\'eziau and Katarzyna Gan-Krzywoszynska (Eds.), 223-242, Philosophia Verlag, Munich.

\bibitem{deRondeMassri14} de Ronde, C. and Massri, C., 2014, ``Revisiting the First Postulate of Quantum Mechanics: Invariance and Physically Reality", {\it Philosophy of Science Archive}, http://philsci-archive.pitt.edu/11189/

\bibitem{Dieks88b} Dieks, D., 1988, ``Quantum Mechanics and
Realism'', {\it Conceptus XXII}, {\bf 57}, 31-47.

\bibitem{Dieks10} Dieks, D., 2010, ``Quantum Mechanics, Chance and Modality", {\it Philosophica}, {\bf 83}, 117-137.

\bibitem{Dirac74} Dirac, P. A. M., 1974, {\it The Principles of Quantum Mechanics}, 4th Edition, Oxford University Press, London.

\bibitem{EPR} Einstein, A., Podolsky, B. and Rosen, N., 1935,
``Can Quantum-Mechanical Description be Considered Complete?'', {\it
Physical Review}, {\bf 47}, 777-780.

\bibitem{Folse87} Folse, H. J., 1987, ``Niels Bohr's Concept of Reality'',
{\it Symposium on the foundations of Modern Physics 1987}, 161-179,
P. Lathi and P. Mittelslaedt (Eds.) World Scientific, Singapore.

\bibitem{FRD12}  Freytes, H., de Ronde, C. and Domenech, G., 2012, ``The square of opposition in orthodmodular logic'', in {\it Around and Beyond the Square of Opposition: Studies in Universal Logic}, Jean-Yves B\'eziau and Dale Jacquette (Eds.), pp. 193-201, Springer, Basel.

\bibitem{Friedman00} Friedman, M., 2000, {\it A Parting of the Ways:
Carnap, Cassirer, and Heidegger}, Open Court, Chicago.

\bibitem{FuchsPeres00} Fuchs, C. and Peres, A., 2000, ``Quantum theory
needs no `interpretation' '', {\it Physics Today}, {\bf 53}, 70.

\bibitem{Heis58} Heisenberg, W., 1958, {\it Physics and Philosophy},
World perspectives, George Allen and Unwin Ltd., London.

\bibitem{HilgevoordUffink01} Hilgevoord, J. and Uffink, J., 2001, ``The
Uncertainty Principle", {\it The Stanford Encyclopedia of Philosophy
(Winter 2001 Edition)}, E. N. Zalta (Ed.), URL:
http://plato.stanford.edu/archives/win2001/entries/qt-uncertainty/.

\bibitem{Howard93} Howard, D., 1993, ``Was Einstein Really a Realist?'',
{\it Perspectives on Science}, {\bf 1}, 204-251.

\bibitem{Howard94} Howard, D., 1994, ``Einstein, Kant, and the
Origins of Logical Empiricism'', In {\it Logic, Language, and the
Structure of Scientific Theories: Proceedings of the
Carnap-Reichenbach Centennial, University of Konstanz, 21-24 May
1991}, 45-105, W. Salmon and G. Wolters (Eds.), University of
Pittsburgh Press, Pittsburgh.

\bibitem{Kauark} Kauark-Leite, P., 2004, {\it The transcendental
approach and the problem of language and reality in quantum
mechanics}, PhD thesis, Centre de Recherche en Epist\'{e}mologie
Appliqu\'{e}e - \'{E}cole Polytechnique.

\bibitem{KS} Kochen, S. and Specker, E., 1967, ``On the problem
of Hidden Variables in Quantum Mechanics", {\it Journal of
Mathematics and Mechanics}, {\bf 17}, 59-87. Reprinted in Hooker,
1975, 293-328.

\bibitem{Lahti80} Lahti, P., 1980, ``Uncertainty and Complementarity
in Axiomatic Quantum Mechanics", {\it International Journal of
Theoretical Physics}, {\bf 19}, 789-842.

\bibitem{Meillassoux} Meillassoux, Q., 2006, {\it Apr\`es la
finitude. Essai sur la n\'ecessit\'e de la contingence}, \'Editions
du Seuil, 2006.

\bibitem{Nitiniluoto87} Niniluoto, I., 1987, ``Varities of Realism",
In {\it Symposium on the foundations of Modern Physics 1987},
459-483, P. Lathi and P. Mittelslaedt (Eds.),  World Scientific,
Singapore.

\bibitem{Pauli94} Pauli, W., 1994, {\it Writings on Physics and
Philosophy}, Enz, C. and von Meyenn, K. (Eds.), Springer-Verlag,
Berlin.

\bibitem{Petersen63} Petersen, A., 1963, ``The Philosophy of Niels
Bohr'', {\it Bulletin of the Atomic Scientists}, Sep 1963, 8-14.

\bibitem{Plotnyski} Plotnyski, A., 1994, {\it Complementarity: Anti-Epistemology
After Bohr and Derrida}, Duke University Press, Durham.

\bibitem{Priest87} Priest, G., 1987, {\it In Contradiction}. Nijhoff, Dordrecht.

\bibitem{Redei01} R\'{e}dei, M., 2001, ``Von Neumann's concept of quantum logic and quantum
probability'', in {\it John von Neumann and the Foundations of
Quantum Physics}, M. R\'{e}dei and M. St\"{o}tzner (Eds.), 153-172,
Kluwer Academic Publishers, Dordrecht.

\bibitem{RedeiSummers07} R\'{e}dei, M. and Summers, S.J., 2007, ``Quantum probability
theory'', {\it Studies in the History and Philosophy of Modern Physics},
\textbf{38}, 390-417.

\bibitem{Scavino00} Scavino, D., 2001, {\it La filosof\'ia actual}, Paidos,
Buenos Aires.

\bibitem{Schr35} Schr\"odinger, E., 1935, ``The Present Situation
in Quantum Mechanics'', {\it Naturwiss}, {\bf 23}, 807. Translated
to english in {\it Quantum Theory and Measurement}, J. A. Wheeler
and W. H. Zurek (Eds.), 1983, Princeton University Press, Princeton.

\bibitem{Smets05} Smets, S., 2005,  ``The Modes of Physical Properties
in the Logical Foundations of Physics'', {\it Logic and Logical
Philosophy}, {\bf 14}, 37-53.

\bibitem{Vaidman08} Vaidman, L., 2008, ``Many-Worlds Interpretation of Quantum Mechanics", In {\it The Stanford Encyclopedia of Philosophy (Fall 2008 Edition)}, E. N.
Zalta (Ed.), URL:
http://plato.stanford.edu/archives/fall2008/entries/qm-manyworlds/.

\bibitem{VF91} Van Fraassen, B. C., 1991, {\it Quantum Mechanics: An
Empiricist View}, Clarendon, Oxford.

\bibitem{CFVW} von Weizs¬acker, C.F.,  1955, ``Komplementarit\"at und Natuurwissenschaft'', {\it Die Natuurwisenchaften}, p. 42, n. 19-20. Translated as ``Complementariedad y L\'ogica'' in {\it La Imagen F\'isica del Mundo}, 1974, Biblioteca de Autores Crsitianos, Madrid.

\bibitem{WZ} Wheeler, J. A. and Zurek, W. H. (Eds.) 1983, {\it Theory and
Measurement}, Princeton University Press, Princeton.

\end{thebibliography}

\newtheorem{theo}{Theorem}[section]

\newtheorem{definition}[theo]{Definition}

\newtheorem{lem}[theo]{Lemma}

\newtheorem{met}[theo]{Method}

\newtheorem{prop}[theo]{Proposition}

\newtheorem{coro}[theo]{Corollary}

\newtheorem{exam}[theo]{Example}

\newtheorem{rema}[theo]{Remark}{\hspace*{4mm}}

\newtheorem{example}[theo]{Example}

\newcommand{\proof}{\noindent {\em Proof:\/}{\hspace*{4mm}}}

\newcommand{\qed}{\hfill$\Box$}

\newcommand{\ninv}{\mathord{\sim}} 

\newtheorem{postulate}[theo]{Postulate}

\section{Interpreting Quantum Mechanics}

Regarding its formal structure we could say that quantum mechanics (QM) seems to be a ``finished theory''. In terms of empirical adequacy, it provides outstanding results, its mathematical structure ---developed in the first three decades of the 20th century by Werner Heisenberg, Pascual Jordan, Max Born, Erwin Schr\"odinger and Paul Dirac--- seems able to predict any experiment we can think of. However, apart from its fantastic accuracy, even today, more than one century after its creation, its physical interpretation remains an open problem. In the standard formulation, QM assigns a quantum mechanical state to a system, but `the state' has a meaning only in terms of the outcomes of the measurements performed and not in terms of `something' which one can coherently relate to physical reality \cite{deRondeMassri14}. It is not at all clear, apart from measurement outcomes, what should be the interpretation of a vector in Hilbert space, in particular, and of the formal structure, in general. If we are to ask too many questions regarding the physical meaning of the theory, problems start to pop up and simple answers seem doomed to inconsistency.

From the very beginning of the voyage, the problem of the founding fathers was to find a picture (an \emph{anschauliche} content), a physical representation which would allow to explain {\it what} QM was talking about. This idea guided the early attempt of Luis de Broglie with his matter-wave theory in 1924, and of Schr\"odinger with the introduction of his wave equation in 1926. It is also well known that Albert Einstein was very uncomfortable with the unclear reference of QM to physical reality. His concerns were matter of debate in the 1927 Solvay conference in Brussels and also, later on, his criticisms were exposed in the famous EPR paper written in 1935 \cite{EPR}. However, very soon, it became clear that the interpretation of QM faced deep problems when attempting to provide a coherent description of physical reality. Due to this impossibility, very soon more pragmatic stances were developed. For example, Paul Dirac \cite[p. 10]{Dirac74} writes in his famous book, {\it The Principles of Quantum Mechanics}:  ``[...] the main object of physical science is not the provision of physical pictures, but is the formulation of laws governing phenomena and the application of these laws to the discovery of new phenomena. If a picture exists, so much the better; but whether a picture exists or not is a matter of only secondary importance." Taking distance from the physical representation of the formalism, Niels Bohr, one of the key figures in the creation and development of QM, developed a scheme in which he restricted the reference of the theory to classical phenomena. His choice, in resonance with the philosophical movements of the period, determined the problems and questioning of the future generations of physicists and philosophers of physics working on the fundamental questions about quantum theory.

\section{Niels Bohr and the Linguistic Turn in Physics}

Niels Bohr might have been the most influential figure who tried ---and succeeded to great extent--- to expel metaphysical questions from the debate regarding QM. As remarked by Arthur Fine: 

\begin{quotation}
\noindent {\small ``These instrumentalist moves, away from a realist construal of the emerging quantum theory, were given particular force by Bohr's so-called `philosophy of complementarity'; and this nonrealist position was consolidated at the time of the famous Solvay conference, in October of 1927, and is firmly in place today. Such quantum nonrealism is part of what every graduate physicist learns and practices. It is the conceptual backdrop to all the brilliant success in atomic, nuclear, and particle physics over the past fifty years. Physicists have learned to think about their theory in a highly nonrealist way, and doing just that has brought about the most marvelous predictive success in the history of science.'' \cite[p. 1195]{PS}}
\end{quotation}

\noindent At distance from anti-metaphysical construals Einstein was a strong defender of physical representation. As recalled by Wolfgang Pauli:

\begin{quotation}
\noindent {\small ``{\it {\small Einstein}}'s opposition to [QM] is again reflected in his papers which he published, at first in collaboration with {\small \emph{Rosen}} and {\small \emph{Podolsky}}, and later alone, as a critique of the concept of
reality in quantum mechanics. We often discussed these questions
together, and I invariably profited very greatly even when I could
not agree with {\small \emph{Einstein}}'s view. `Physics is after
all the description of reality' he said to me, continuing, with a
sarcastic glance in my direction `or should I perhaps say physics is
the description of what one merely imagines?' This question clearly
shows {\small \emph{Einstein}}'s concern that the objective
character of physics might be lost through a theory of the type of
quantum mechanics, in that as a consequence of a wider conception of
the objectivity of an explanation of nature the difference between
physical reality and dream or hallucination might become blurred.'' \cite[p. 122]{Pauli94}}
\end{quotation}

It is important to remark that the debate which took place between Einstein and Bohr regarding physical reality can be only understood as part of the neo-Kantian tradition and discussion which was taking place in German speaking countries at the end of the 19th and beginning of the 20th century. This discussion is very well exposed by Michael Friedman in his beautiful book, {\it A Parting of the Ways} \cite{Friedman00}. From this perspective, both Einstein and Bohr were discussing from within representation, considering specifically the conditions of possibility to access phenomena. Despite the account provided by many, Einstein and Bohr were not part of the ---extensively addressed in the philosophy of science literature--- realist anti-realist debate (see for discussion \cite{Howard93, Howard94}).  

According to our reading of Bohr,\footnote{We acknowledge there are almost as many interpretations of Bohr as physicists and philosophers of science. Even though the orthodoxy has been to interpret Bohr from a neo-Kantian perspective, there are also ontological interpretations of Bohr such as those proposed by Folse \cite{Folse87} and Dieks \cite{Dieks88b}. In particular, Dieks interprets complementarity as an ontological notion which relates `experimental situations'.} the preeminence of language within his own philosophical scheme can be only understood in relation to the bigger philosophical movement which was taking place in Europe and has been called the ``linguistic turn''.\footnote{We take the ling\"uistic turn to be a moment with multiple lines of philosophical investigation which can be comprised by the importance of language as a fundament. See the interesting analysis of Dardo Scavino in \cite{Scavino00}.} So even though QM arose in relation to the criticism of Kant's epistemology ---escaping from the domains of the {\it a priori} categories and forms of intuition---, it was soon {\it re-covered} by the neo-Kantian scheme of thought. It was the Danish physicist who was capable of introducing physics into the philosophical movement of the linguistic turn, shifting quantum theory from ontological concerns into epistemological ones. Niels Bohr was the first to initiate the re-turn of physics back into the domain of philosophy after the revolution produced by relativity theory and QM. Just like Immanuel Kant did with Newtonian mechanics, turning upside down the relation of power between physics and philosophy (see \cite{Meillassoux}), Bohr was able to constrain the strength of physics within the limitations imposed by that which would now play the role of the {\it a priori}: classical language.\footnote{It is then not a surprise to notice that the philosophy of Niels Bohr has been directly engaged with philosophers like Ludwig Wittgentstein (see \cite{Nitiniluoto87}), Jaques Derrida (see \cite{Plotnyski}), and of course, Immanuel Kant (see \cite{Kauark} for an extensive review of the relation between Bohr and Kant's philosophy).}

Niels Bohr's ideas have played a central role in the development of physics in the 20th century, placing the discipline within the main philosophical line of discussion of the period, namely, the problem of language and its relation to ontology and epistemology. The linguistic turn is a technical term in the history of philosophy according to which all problems in philosophy are problems of language. We do not claim that Bohr knew that movement or was explicitly part of it. Rather, we point to the fact that, quite independently of this movement, Bohr took for himself many of the discussions and problems involved within such philosophical stance. A clear statement regarding this point is the famous quotation by Aage Petersen. According to the long time assistant of Bohr, when asked whether the quantum theory could be considered as somehow mirroring an underlying quantum reality Bohr \cite[p. 8]{WZ} declared the following: ``There is no quantum world. There is only an abstract quantum physical description. It is wrong to think that the task of physics is to find out how nature is. Physics concerns what we can say about nature" The problem was then how to secure {\it communication}. Instead of going directly back into Kant's {\it a priori}, Bohr made a detour into the realm of language, finding his cornerstone, his ``clear and distinct idea'' in the the language used by classical physics.

\begin{quotation}
\noindent {\small ``Even when the phenomena transcend the scope of classical
physical theories, the account of the experimental arrangement and the recording of observations must be given in plain language, suitably supplemented by technical physical terminology. This is a clear logical demand, since the very word {\small {\it experiment}} refers to a situation where we can tell others what we have done and
what we have learned." \cite[p. 7]{WZ}}
\end{quotation}

\noindent Making his point even more explicit Bohr \cite[p. 7]{WZ} claimed that: ``[...] the unambiguous interpretation  of any measurement must be essentially framed in terms of classical physical theories, and we may say that in this sense the language of Newton and Maxwell
will remain the language of physicists for all time.'' In this same sense, Bohr seemed to take distance from ontological questioning. As also noted by Aage Petersen:

\begin{quotation}
\noindent {\small ``Traditional philosophy has accustomed us to regard
language as something secondary and reality as something primary.
Bohr considered this attitude toward the relation between language
and reality inappropriate. When one said to him that it cannot be
language which is fundamental, but that it must be reality which, so
to speak, lies beneath language, and of which language is a picture,
he would reply, ``We are suspended in language in such a way that we
cannot say what is up and what is down. The word `reality' is also a
word, a word which we must learn to use correctly'' Bohr was not
puzzled by ontological problems or by questions as to how concepts
are related to reality. Such questions seemed sterile to him. He saw
the problem of knowledge in a different light." \cite[p. 11]{Petersen63}}
\end{quotation}

\noindent Bohr's characterization of physics goes then together with his
linguistically based pragmatic account:

\begin{quotation}
\noindent {\small ``Physics is to be regarded not so much as the study of
something a priori given, but rather as {\small{\it the development
of methods of ordering and surveying human experience.}} In this
respect our task must be to account for such experience in a manner
independent of individual subjective judgement and therefor
{\small{\it objective in the sense that it can be unambiguously
communicated in ordinary human language.}}" 
\cite{Bohr60} (emphasis added)} \end{quotation}

However, this line of thought ---which considers a physical
theory in pragmatic and linguistic terms--- might be regarded as ending up in a path which seems difficult to maintain, at least in the case we are still willing to state that there is something like ``physical reality'' of which our theories talk about.
In a spirit very similar to the ideas expressed by Dirac, and after
endless discussions regarding the meaning of the quantum, in the
year 2000, exactly one century after the beginning of the voyage,
Christopher Fuchs and Asher Peres finally took this line to its
unavoidable conclusion in a paper entitled: {\it Quantum Theory
Needs no `Interpretation'}. There, they wrote:

\begin{quotation}
\noindent {\small ``[...] quantum theory does not describe physical reality.
What it does is provide an algorithm for computing probabilities for
the macroscopic events (``detector clicks") that are the
consequences of experimental interventions. This strict definition
of the scope of quantum theory is the only interpretation ever
needed, whether by experimenters or theorists." \cite[p. 1]{FuchsPeres00}}
\end{quotation}

\noindent Of course this emphasis on prediction to the detriment of
description can be severely questioned. The main objection against
this instrumentalistic point of view is that the success of a theory
can not be explained, that is to say, we do not know how and why
quantum physics is in general able to carry out predictions (and in
particular with such a fantastic accuracy). Or in other words, there is no physical representation of what is going on according to the theory. Undoubtedly a ``hard"
instrumentalist may simply refuse to look for such an explanation,
since it is in fact the mere effectiveness of a theory which
justifies it, so that he may not be interested in advancing towards
a justification of that effectiveness. If one takes such a position
there is nothing left to say. Just like the Oracle of Delphi
provided always the right answer to the ancient Greeks, QM provides us with the correct probability distribution for every experiment we can think of.

\section{The Re-Turn of Classical Metaphysics}

Since the second world war the philosophical analysis of science, and of quantum theory in particular, has been an almost exclusive field owned by analytic philosophy. Although the analytical tradition is an inheritor ---via logical positivism and logical empiricism--- of a deep criticism to metaphysics, strangely enough, within analytic philosophy of physics the attempt to return to a classical metaphysical scheme of thought seems to be a recursive element, specially in the philosophy of QM during the second half of the 20th century. 

Indeed, the position of Bohr, which can be very well regarded in close continuation to analytic
concerns against metaphysics, was replaced after the war by approaches to QM, such as, for example, Bohmian mechanics and DeWitt's many worlds interpretation, which recovered the classical metaphysical foundation of thought and understanding. While Bohr attempted to analyze the logical structure of the theory and concentrated on the analysis of phenomena, these new lines of research intended ``to restore a classical way of thinking about {\it what there is}.'' \cite[p. 74]{Bacciagaluppi96} It seems in this case a bit ironic that the aversion professed by many philosophers of physics ---which are part of the analytic tradition--- to Bohr's ideas does not recognize the profound connection of his thought to analytic philosophy itself. These same philosophers choose ---knowingly or not--- for metaphysical schemes going against their own tradition. In the case of many worlds interpretation the metaphysical step goes as far as to propose non-observable worlds in order to explain the formal aspects of QM. Also, from a metaphysical point of view, the many worlds attempt seems to end up in an extreme violation of Ockham's principle: ``Entities are not to be multiplied beyond necessity".\footnote{Although Lev Vaidman \cite{Vaidman08} claims that: ``in judging physical theories one could reasonably argue that one should not multiply physical laws beyond necessity either (such a verion of Ockham's Razor has been applied in the past), and in this respect the many worlds interpretations is the most economical theory. Indeed, it has all the laws of the standard quantum theory, but without the collapse postulate, the most problematic of physical laws." One could argue however, that due to the existence of modal interpretations, which are also no collapse interpretations and share the same formal structure as many worlds, there is no clear argument why one should be forced into this expensive metaphysical extension.} In the case of Bohmian mechanics the metaphysical presuppositions got as far as postulating that QM must talk about classical particles with definite trajectories. Bitbol notices in this respect \cite[p. 8]{Bitbol10} that: ``Bohm's original theory of 1952 is likely to be the most metaphysical (in the strongest, speculative, sense) of all readings of QM. It posits free particle trajectories in space-time, that are unobservable in virtue of the theory itself." Furthermore, that which should play the role of space-time in the mathematical formalism varies its dimension with the addition or substraction of particles breaking the initial attempt to recover trajectories in space-time. It is not at all clear that these kind of attempts bring more solutions than problems. 

The Danish physicist remained agnostic regarding the metaphysical concerns raised especially by Einstein, but also those of Heisenberg and Pauli. He tried by all means to restrict his analysis to the empirical data as exposed by classical physical theories and language, and not go beyond the interpretation of the formalism in terms of a new conceptual scheme. According to Bohr \cite[p. 7]{WZ} ``it would be a misconception to believe that the difficulties of the atomic theory may be evaded by eventually replacing the concepts of classical physics by new conceptual forms.'' Contrary to this approach, Bohmian mechanics and many worlds interpretations, two of the most important interpretational lines of research in present philosophy of QM today compose their analysis with heavy metaphysical commitments based to great extent in the actualist picture put forward by Newtonian physics (see for discussion \cite{RFD14b}). Rather than starting from the analysis of the formal structure of the theory, the metaphysical presuppositions constitute the very foundation and center of gravity of such interpretations. Some of these interpretations even attempt in some cases (e.g., Bohmian mechanics and GRW theory) to change the formalism in order to recover ---at least some of--- our classical (metaphysical) conception of the world.

\section{Complementarity and Paraconsistency}

Paraconsistent Logics (PL) are the logics of inconsistent but non-trivial
theories. The origins of PL go back to the first systematic studies
dealing with the possibility of rejecting the PNC. PL was
elaborated, independently, by Stanislaw Jaskowski in Poland, and by
Newton da Costa in Brazil, around the middle of the last century (on
PL, see, for example: \cite{daCostaKrauseBueno07}). A theory $T$
founded on the logic $L$, which contains a symbol for negation, is
called inconsistent if it has among its theorems a sentence $A$ and
its negation $\neg A$; otherwise, it is said to be consistent. $T$
is called trivial if any sentence of its language is also a theorem
of $T$; otherwise, $T$ is said to be non-trivial. In classical
logics and in most usual logics, a theory is inconsistent if, and
only if, it is trivial. $L$ is paraconsistent when it can be the
underlying logic of inconsistent but non trivial theories. Clearly,
no classical logic is paraconsistent. 

In the context of QM, da Costa and Krause have put forward \cite{daCostaKrause06} a PL in order to provide a suitable formal scheme to consider the notion of complementarity introduced by Bohr in 1927 during his famous `Como Lecture'.\footnote{The logical understanding of complementarity has an interesting history which goes back to Carl Friedrich von Weizs\"acker who wrote an article named: ``Komplementarit\"at und Natuurwissenschaft'' \cite{CFVW} for Bohr's 70th birthday. In this article he explained the concept of complementarity in terms of {\it parallel complementarity} and {\it circular complementarity}. The difficulties to understand complementarity is exposed by a rectification added at the end of the same paper in which von Weizs\"acker explains that he received a letter from Bohr expressing that complementarity can be only defined with respect to phenomena, and as the Schr\"odinger wave equation is just an abstract magnitude of calculus and it does not designate in itself any phenomena, such circular complementarity is by no means possible and only parallel complementarity should be taken into account.} The notion of complementarity was developed in order to consider the contradictory classical representations found in the double-slit experiment; i.e., the representation provided by the notions of `wave' and `particle'. According to Bohr \cite[p. 103]{daCostaKrause06}: ``We must, in general, be prepared to accept the fact that a complete elucidation of one and the same object may require diverse points of view which defy a unique description.'' The starting point of analysis was for Bohr the classical description of experimental arrangements univocally represented by classical language (with the aid of physics). Bohr considered the wave-particle duality present in the double-slit experiment as expressing the most important character of quantum theory. The resolution of this duality was provided via his own notion of complementarity. Bohr's agenda was focused in fulfilling the consistency requirements of the quantum formalism to apply the well known classical scheme. In this respect Heisenberg's principle was only considered as providing the limits of certainty and applicability of classical concepts as such. The classical scheme would then remain that which secured the knowledge provided by QM, and analogously, Heisenberg's {\it uncertainty relations} that which secured the knowledge provided by the more general principle of complementarity.\footnote{It is important to notice that Heisenberg's relations can be directly derived from the mathematical scheme of the theory, as a direct consequence of the quantum postulate. Today, we have more elements to make precise the relation between both principles, see for example, the analysis of Pekka Lahti in his thesis \cite{Lahti80}. As remarked by J. Hilgevoord and J. Uffink \cite{HilgevoordUffink01}: ``On the one hand, Bohr was quite enthusiastic about Heisenberg's ideas which seemed to fit wonderfully with his own thinking. Indeed, in his subsequent work, Bohr always presented the uncertainty relations as the symbolic expression of his complementarity viewpoint. On the other hand, he criticized Heisenberg severely for his suggestion that these relations were due to discontinuous changes occurring during a measurement process. Rather, Bohr argued, their proper derivation should start from the indispensability of both particle and wave concepts. He pointed out that the uncertainties in the experiment did not exclusively arise from the discontinuities but also from the fact that in the experiment we need to take into account both the particle theory and the wave theory."} As Leon Rosenfeld makes the point:

\begin{quotation}
{\small ``Bohr wanted to pursue the epistemological analysis one step further [than Heisenberg], and in particular to understand the logical nature of the mutual exclusion of the aspects opposed in the particle-wave dualism. From this point of view the indeterminacy relations appear in a new light. [...] The indeterminacy relations are therefore essential to ensure the consistency of the theory, by assigning the limits within which the use of classical concepts belonging to the two extreme pictures may be applied without contradiction. For this novel logical relationship, which called in Bohr's mind echoes of his philosophical meditations over the duality of our mental activity, he proposed the name `complementarity', conscious that he was here breaking new ground in epistemology." \cite[p. 59]{WZ}} \end{quotation}

In \cite{daCostaKrause06} the proposal was to go further into the notion of {\it complementary theories}.\footnote{This idea of considering complementary theories goes in line with the complementary descriptions approach proposed in \cite{deRonde05}.} 

\begin{quotation}
\noindent {\small ``[...] we shall say that a theory $T$ admits complementarity interpretation, or that $T$ is a $C$-theory, if $T$ encompasses `true' formulas $\alpha$ and $\beta$ (which may stand for Jammer's $D_1$ and $D_2$ respectively) which are `mutually exclusive' in the above sense, for instance, that their conjunction yields to a strict contradiction if classical logic is applied. In other words, if $\vdash$ is the symbol of deduction of classical logic, then, $\alpha$ and $\beta$ being complementary, we have $\alpha$, $\beta \vdash \gamma \wedge \neg \gamma$ for some $\gamma$ of the language of $T$" \cite[p. 111]{daCostaKrause06}} \end{quotation}

The proposal of da Costa and Krause was also considered by Jean-Yves B\'eziau \cite{Beziau14} stressing in his case the notion of {\it viewpoint} and taking into account the Square of Opposition. According to B\'eziau: 

\begin{quotation}
{\small ``In modern physics we have a subcontrary opposition between wave and particle in the sense that the proposition `K is a particle' and `K is a wave' can both be true but cannot both be false. [...]  [O]ne may say that with particle and wave there is an opposition because something cannot be at the same time a particle and a wave, due to the very nature of wave and particle, in the same sense that something cannot be a square and a circle. But why then can we say that `K is a particle' and `K is a wave' can both be true but not that `K is a circle' and `K is a square' can both be true? In fact it is also possible to say that these two geometrical propositions are both true, but from a different perspective, which is not the usual flat one." \cite{Beziau14}} \end{quotation}

B\'eziau then continues to analyze the situation taking into account more explicitly the position of Bohr and his complementarity approach. 

\begin{quotation}
{\small ``[Bohr] argues that there are no direct contradiction: from a certain point of view `K is a particle', from another point of view `K is a wave', but these two contradictory properties appear in different circumstances, different experiments. Someone may ask: what is the absolute reality of K, is K a particle or is K a wave? One maybe has to give away the notion of objective reality." \cite{Beziau14}} \end{quotation}

\noindent At this point it becomes clear that the paraconsistency implied by B\'eziau relates to the perspective assumed by the observer. The contradiction appears through the choice of the subject between the different experimental set-ups. Following this idea he develops a logical theory where the central concept is the concept of viewpoint. The price to pay, as clearly acknowledged by B\'eziau, is to abandon the notion of objective physical reality. In the following section we will argue that paraconsistency can also help us to develop a new idea of physical reality that would allow us to consider QM as a physical theory providing an objective representation of a world outside there.

\section{Quantum Superpositions and Paraconsistency}

In classical physics, every physical system may be described exclusively by means of its \emph{actual properties}, taking `actuality' as expressing the \emph{preexistent} mode of being of the properties themselves, independently of observation ---the `pre' referring to its existence previous to measurement. Each system has a determined state characterized mathematically in terms of a point in phase space. The change of the system may be described by the change of its actual properties. Potential or possible properties are considered as the points to which the system might arrive in a future instant of time. As Dieks makes the point: ``In classical physics the most fundamental description of a physical system (a point in phase space) reflects only the actual, and nothing that is merely possible. It is true that sometimes states involving probabilities occur in classical physics: think of the probability distributions $\rho$ in statistical mechanics. But the occurrence of possibilities in such cases merely reflects our ignorance about what is actual." It is then important to recognize that the main character which makes possible this physical description in terms of actual properties, and more in general, in terms of an actual state of affairs is the fact that the mathematical structure allows a global valuation of all properties \cite{deRondeMassri14}. 

In QM the representation of the state of a system is given by a ray in Hilbert space $H$ and physical magnitudes are represented by operators on $H$ that, in general, do not commute. As a consequence, the Kochen-Specker theorem precludes the possibility of a global valuation of all properties independently of the context \cite{KS}. It then becomes difficult to affirm that all quantum magnitudes are \emph{simultaneously preexistent}. In order to restrict the discourse to different sets of commuting magnitudes, different Complete Sets of Commuting Operators (CSCO) have to be chosen. The choice of a particular representation (given by a CSCO) determines the basis in which the observables diagonalize and in which the ray can be expressed. Thus, the ray can be written as different linear combinations of states:

\begin{equation}\label{cl}
\alpha_{i}|\varphi_{i}^{B1}> + \alpha_{j}|\varphi^{B1}_j> =
|\varphi^{B2}_q> =\beta_{m}|\varphi_{m}^{B3}> + \beta_{n}
|\varphi_{n}^{B3}> + \beta_{o} |\varphi_{o}^{B3}>
\end{equation}

\noindent Each linear combination of states is also called a {\it quantum superposition}. The Born interpretation tells us that the numbers that accompany each state in square modulus compute the probability of finding that particular state. It is also well known that such probability cannot be interpreted in terms of ignorance \cite{Redei01, RedeiSummers07}. As remarked by Schr\"odinger in a letter to Einstein:  

\begin{quotation}
{\small ``It seems to me that the concept of probability [in QM] is terribly mishandled these days. Probability surely has as its substance a statement as to whether something is or is not the case ---of an uncertain statement, to be sure. But nevertheless it has meaning only if one is indeed convinced that the something in question quite definitely is or is not the case. A probabilistic assertion presupposes the full reality of its subject.''  \cite[p. 115]{Bub97}} \end{quotation}

If we consider a typical Stern-Gerlach experiment, given a spin $1/2$ system whose state is $\frac{1}{2^{1/2}} |\uparrow_{x}> + |\downarrow_{x}>$, orthodox QM tells us that we shall obtain either a `click' in the upper part of the screen with probability 0.5 (which relates to the state $|\uparrow_{x}>$), or, a `click' in the bottom part of the screen also with probability 0.5 (which relates to the state $|\downarrow_{x}>$). However, it is not clear at all what {\it is} the state before the measurement since, on the one hand, one cannot claim due to the formalism of the theory that the measurement discovers a preexistent actual reality, and on the other hand, actuality itself precludes the existence of the seemingly exclusive possibilities. According to our interpretation there are good reasons to claim that, in the just mentioned example, both states (`$|\uparrow_{x}>$' and $|`\downarrow_{x}>$') should be regarded as existent ---rather than exclusive possibilities. As a matter of fact, quantum superpositions `evolve' according to the Schr\"odingier equation of motion, `interact' with other superposition states ---creating the famous entanglement of multiple states--- and can `be predicted' according to the rules of QM. But if, within a physical theory there is a mathematical expression which allows us to calculate the evolution of its terms, its interaction with other mathematical expressions of the type and predict its possible results, after or before the interaction, then it becomes reasonable to claim that there is something ``physically real'' about such mathematical expression. 

The problem at this point is that while these two states, $|\uparrow_{x}>$ and $|\downarrow_{x}>$, seem to exist before measurement as somehow contradictory states, the realm of actuality denies ---by definition--- the existence of contradictions. This weird fact about quantum supertpositions was cleverly explained by Schr\"odinger through his famous cat experiment which produced a zombie cat half dead and half alive. At this point one might consider two different interpretative strategies to solve the paradox. The first, to find the way to interpret quantum superpositions  escaping contradictions, forcing the actual realm of existence in order to interpret the formalism ---e.g. as many worlds or the modal interpretation of Dieks attempts to do. The second, more radical path, is to develop a realm of existence which is not that of actuality. It is this latter possibility which we find more interesting.

In \cite{daCostadeRonde13}, Newton da Costa together with the author of this paper argued in favor of the possibility to consider quantum superpositions in terms of a paraconsistent approach. We discussed the idea that, even though most interpretations of QM attempt to escape {\it contradictions}, there are many reasons that indicate it could be worth while to assume that the terms in a quantum superposition can be, in principle, contradictory ---as in the just mentioned example. It should be also stressed that the Paraconsistent Approach to Quantum Superpositions (PAQS) does not present an interpretation of QM but just a way to account for quantum superpositions in terms of paraconsistent logics. This formal approach might lead to different interpretations. However, one may also recognize that intuitively, the approach might favor an interpretation in which superpositions are considered as contradictory physical existents. 

In \cite{ArenhartKrause14a} Arenhart and Krause raised several arguments against the PAQS, one of the main concerns having to do with the consideration of quantum superpositions as contradictory existents. In \cite{deRonde14a}, the author of this paper argued, firstly, that the obstacles presented by Arenhart and Krause are based on a specific metaphysical stance, which we characterized in terms of what we called the Orthodox Line of Research (OLR). Secondly, that this is not necessarily the only possible line, and that a different one, namely, a Constructive Metaphysical Line of Research (CMLR) provides a different perspective in which the PAQS can be regarded as a valuable prospect that could be used by different interpretations of QM. 

It is interesting to point out that the OLR implicitly embraced by Krause and Arenhart reflects the Bohrian perspective towards the problem of interpretation in QM. Indeed, the OLR has always debated within the limits imposed by the Danish physicist ---this has been Bohr's true success. As we have discussed in detail in \cite{deRonde14a} this path can be condensed in two main metaphysical presuppositions which block the conceptual development of quantum superpositions. The first presupposition relates to what Bohr called the {\it correspondence principle}, an idea which has been later on reconsidered in the literature in terms of what is known to be the quantum to classical limit \cite{BokulichCP}.

\begin{enumerate}
\item {\bf Quantum to Classical Limit:}  The principle that one can find a  bridge between classical mechanics and QM; i.e., that the main notions of classical physics can be used in order to explain quantum theory.
\end{enumerate}

\noindent The second metaphysical principle can be also traced back to Bohr's claim that physical experience needs to be expressed exclusively in terms of classical language \cite{BokulichBokulich}. As a matter of fact, if one considers the core of the classical physical and metaphysical representation of the world one is then stuck with two main concepts: `entity' and `actuality' (as a mode of existence). In QM one can also encounter these metaphysical notions as basic elements of any interpretation.

\begin{enumerate}
\item[2.] {\bf Quantum Systems and Actuality:} The principle that one needs to presuppose the metaphysics of entities together with the mode of being of actuality in any interpretation of QM.
\end{enumerate}

The idea proposed in \cite{daCostadeRonde13} to introduce paraconsistency in order to account for quantum superpositions might be regarded as opening the possibility to supplement such formal scheme with new non-classical physical notions. It should be also clear, however, that this is not necessarily the only option, since the PAQS might be regarded as only a formal proposal. But in case we would be willing to step into the realm of metaphysics and advance with an interpretation that accepts {\it contradictory existents}, then we must also come up with a realm of existence different from actuality, breaking the equation that has ruled physics since Newton: `Actuality = Reality'. One such proposal has been put forward in \cite{deRonde13b}. 

As we have argued extensively in \cite{deRonde14a}, if we now go into the problem of interpretation, the idea presented by the PAQS can be supported by the CMLR which imposes a different perspective towards QM, its problems and the questions that need to be answered. Taking into account the need to provide a coherent physical interpretation of QM, our CMLR is based on three main presuppositions:\footnote{This proposal was put forward and discussed in \cite[pp. 56-57]{deRondePHD}. See also \cite{deRonde14YQM}.}

\begin{enumerate}
\item {\bf Closed Representational Stance:} Each physical theory is closed under its own formal and conceptual structure providing access to a specific set of phenomena. The theory provides the constraints to consider, explain and understand physical phenomena.

\item {\bf Formalism and Empirical Adequacy:} The formalism of QM is able to provide (outstanding) empirically adequate results. Empirical adequacy determines the success of a theory and not its (metaphysical) commitment to a certain presupposed conception of the world. The problem is not to find a new mathematical scheme, on the contrary, the `road signs' point in the direction that {\it we must stay close to the orthodox quantum formalism}.

\item {\bf Constructive Stance:} To learn about what the formalism of QM is telling us about reality we might be in need of {\it creating new physical concepts}.
\end{enumerate}

\noindent What is needed according to the CMLR is a radical inversion of orthodoxy and its problems. According to this inversion, for example, the question of contextuality is not a problem which we need to escape but rather a central characteristic that any interpretation of QM should respect. Also the non-separable character of QM, its specificity with respect to identity and individuality, its indeterminate and indeterministic aspects, etc., should be all considered ---rather than obstacles--- as the main elements that can guide us in the development of a coherent interpretation of the theory. Going back to the meaning of quantum superpositions, instead of considering {\it the measurement problem} we should focus instead on the analysis of what we have called {\it the superposition problem} (see for discussion \cite{deRonde14a}). The constructive stance assumes the radical possibility to consider a different mode of existence. Elsewhere \cite{deRondePHD, deRonde12, deRonde13b}, we have put forward a notion of {\it ontological potentiality} which allows us to discuss an independent realm of existence which ---contrary to the orthodox approach to potentiality--- cannot be reduced to actuality. Indeed, the PAQS, properly supplemented by the CMLR, allows us to interpret quantum superpositions in terms of potential contradictory existents.

\section{Epistemological and Ontological Contradictions}

From the above discussions, it is interesting to notice that the use of paraconsistency in QM can be considered from two very distinct perspectives, according to the philosophical stance that one assumes in order to approach interpretational issues. On the one hand, we have the paraconsistency implied by the epistemological approach to QM which deals with the ``contradiction'' between multiple classical representations. As remarked by da Costa and Krause:

\begin{quotation}
{\small ``It should be emphasized that our way of characterizing complementarity
does not mean that complementary propositions are always contradictory, for
$\alpha$ and $\beta$ above are not necessarily one the negation of the other. However,
as complementary propositions, we may derive from them (in classical logic) a
contradiction; to exemplify, we remark that `$x$ is a particle' is not the direct
negation of `$x$ is a wave', but `$x$ is a particle' {\it entails} that $x$ is not a wave. This
reading of complementarity as not indicating strict contradiction, as we have
already made clear, is in accordance with Bohr himself [...]" \cite{daCostaKrause06}}\end{quotation}
 
\noindent This could be understood as an {\it epistemological contradiction} since it deals with different perspectives or, in terms of Beziau, ``complementary viewpoints''. On the other hand, the paraconsistency discussed within the PAQS implies a more radical stance regarding the meaning of contradictions. Indeed, as we have mentioned above, quantum superpositions ---according to the PAQS--- seem to open the door to consider ``contradictory existents''. Thus one might argue that the PAQS attempts to put forward the introduction of {\it ontological contradictions}.

As remarked by Aristotle:  ``It is impossible for the same thing to belong and not to belong at the same time to the same thing and in the same respect.'' [Metaph. IV 3 1005b19Ð20] On the one hand, it seems that this is not the case of epistemological contradictions which deal with different, rather than contradictory, representations. It makes no sense to say that ``a wave is contradictory to a particle''. On the other hand, ontological contradictions do talk about the same property in the same respect and at the same time. If we consider our earlier Stern-Gerlach experiment the state of affairs is described by the following quantum superposition: $\frac{1}{2^{1/2}} |\uparrow_x \rangle + |\downarrow_x \rangle$, which includes the propositions: `spin up in the x direction' and `spin down in the x direction'.  Both propositions make reference to the same property of spin. 

The introduction of the realm of potentiality reconfigures in itself the meaning of contradiction which has been always considered in terms of the actual realm. Indeed, once we accept the idea that we can have existents in the potential realm we can go further and understand contradictions as related to the path between the potential and the actual realms. If we now take into account the square of opposition we must consider the following set of definitions: 

\begin{enumerate}
\item[]{\bf Contradiction Propositions:} $\alpha$ and $\beta$ are {\it contradictory} when both cannot be true and both cannot be false.
\item[]{\bf Contrariety Propositions:} $\alpha$ and $\beta$ are {\it contrary} when both cannot be true, but both can be false.
\item[]{\bf Subcontrariety Propositions:} $\alpha$ and $\beta$ are {\it subcontraries} when both can be true, but both cannot be false.
\item[]{\bf Subaltern Propositions:} $\alpha$ is {\it subaltern} to proposition $\beta$ if the truth of $\beta$ implies the truth of $\alpha$.
\end{enumerate}

The idea that potentiality determines a contradictory realm goes back to Aristotle himself who claimed that contradictions find themselves in potentiality. Of course, as remarked by Arenhart and Krause, the square of opposition is discussing about actual truth and falsehood. Thus, potentiality is not considered in terms of a mode of existence but rather as mere logical possibility. The introduction of an ontological realm of potentiality changes things drastically. The interesting question is if our representation of quantum superpositions in terms of powers is compatible with the square. We believe it easy to see that such is the case provided special attention is given to the realms involved in the discussion. 

Truth and falsehood are considered in relation to actuality, since in the orthodox view this is the exclusive realm imposing the limits of what can be understood as real. Contrary to the actualist scheme, our notion of ontological potentiality is completely independent of actuality, and thus it makes perfect sense to extend `truth' and `falsity' to this mode of being. We have investigated this possibility in \cite{deRonde14b}. Our redefinition of truth and falsehood with respect to potentiality escapes any subjective choice and regains an objective description of physical reality. The price to pay is to abandon the idea that everything needs to be defined in terms of the actual realm. 

Consider we have a Stern-Gerlach apparatus placed in the $x$ direction, if we have the following quantum superposition: $\alpha  \ | \uparrow_x > +  \  \beta \ | \downarrow_x >$, this means we have `spin up in the $x$ direction, $| \uparrow_x >< \uparrow_x |$, with probability $\alpha$' and `spin down in the $x$ direction, $| \downarrow_x >< \downarrow_x |$, with probability $\beta$' which can be actualized. Is it contradiction or contrariety the best notion suited to account for such existent possibilities\footnote{We have developed an interpretation in which such existent possibilities are discussed in terms of the notion of {\it power}. See \cite{deRonde12, deRonde13b}.} in this quantum experiment? Given this quantum superposition, it is clear that both actualizations of the powers (elementary processes) `$| \uparrow_x >< \uparrow_x |$' and `$| \downarrow_x >< \downarrow_x |$' {\it cannot be simultaneously `true' in actuality}, since only one of them will become actual; it is also the case that both actualizations of the powers (elementary processes) `$| \uparrow_x >< \uparrow_x |$' and `$| \downarrow_x >< \downarrow_x |$' {\it cannot be simultaneously `false' in actuality}, since when we measure the quantum superposition we know we will obtain either the elementary process `spin up in the $x$ direction', `$| \uparrow_x >< \uparrow_x |$', or the  elementary process `spin down in the $x$ direction', `$| \downarrow_x >< \downarrow_x |$'. As we know, given a measurement on the quantum superposition, $\alpha  \ | \uparrow_x > +  \  \beta \ | \downarrow_x >$,  one of the two terms will become actual (true) while the other term will not be actual (false), which implies that {\it both cannot be false}.

\section{Final Remarks}

In this paper we have discussed different approaches regarding the meaning and use of paraconsistency in QM. We have shown how this subject, from very different perspectives, both epistemological and ontological, can provide an interesting discussion and development of the theory. We hope to continue this analysis in future papers.

\section*{Acknowledgements} 

The author would like to thank Newton da Costa, Jean-Yves B\'eziau, Graciela Domenech, Hector Freytes, D\'ecio Krause and Jonas Arenhart for many discussions regarding the subject of paraconsistency in quantum mechanics. This work was partially supported by the following grants: FWO project G.0405.08 and FWO-research community W0.030.06. CONICET RES. 4541-12 (2013-2014).

\end{document}